\documentclass[pre,eqsecnum,preprint,showkeys,preprintnumbers,amssymb,amsfonts]{revtex4}

\input epsf

\usepackage{graphicx}
\usepackage{bm}

\newcommand{\beq}{\begin{equation}}
\newcommand{\eeq}{\end{equation}}
\newcommand{\beqs}{\begin{eqnarray}}
\newcommand{\eeqs}{\end{eqnarray}}

\newtheorem{lemma}{Lemma}[section]
\newtheorem{defi}{Definition}[section]

\newtheorem{propo}{Proposition}[section]

\begin{document}

\title{Acyclic orientations on the Sierpinski gasket}

\author{Shu-Chiuan Chang$^a$}
\email{scchang@mail.ncku.edu.tw} 

\affiliation{(a) \ Department of Physics \\
National Cheng Kung University \\
Tainan 70101, Taiwan}

\begin{abstract}
We study the number of acyclic orientations on the generalized two-dimensional Sierpinski gasket $SG_{2,b}(n)$ at stage $n$ with $b$ equal to two and three, and determine the asymptotic behaviors. We also derive upper bounds for the asymptotic growth constants for $SG_{2,b}$ and $d$-dimensional Sierpinski gasket $SG_d$.
\end{abstract}


\maketitle

\section{Introduction}
\label{sectionI}

The enumeration of the number of acyclic orientations $N_{AO}(G)$ on a graph $G$ is a problem of interest in mathematics \cite{bollobas,reidys,gebhard,gioan} and computer science \cite{kralovic,arantes}. It is well known that the number of acyclic orientations is given by the Tutte polynomial $T(G,x,y)$ evaluated at $x=2$, $y=0$ \cite{welsh} or the chromatic polynomial $P(G,q)$, equivalently the partition function of zero-temperature $q$-state Potts antiferromagnet in statistical mechanics, evaluated at $q=-1$ multiplied by $(-1)^{v(G)}$ where $v(G)$ is the number of vertices \cite{stanley}. It is of interest to consider acyclic orientations on self-similar fractal lattices which have scaling invariance rather than translational invariance. Fractals are geometric structures of (generally noninteger) Hausdorff dimension realized by repeated construction of an elementary shape on progressively smaller length scales \cite{mandelbrot,Falconer}. A well-known example of a fractal is the Sierpinski gasket. We shall derive the recursion relations for the numbers of acyclic orientations on the two-dimensional Sierpinski gasket and its generalization, and determine their asymptotic growth constants defined below. 

\section{Preliminaries}
\label{sectionII}

We first recall some relevant definitions for acyclic orientations and the Sierpinski gasket in this section. A connected graph (without loops) $G=(V,E)$ is defined by its vertex (site) and edge (bond) sets, $V$ and $E$ respectively \cite{bbook,fh}.  Let $v(G)=|V|$ be the number of vertices and $e(G)=|E|$ the number of edges in $G$.  For each edge $e_{ij} = \{v_i,v_j\}$ of $G$, an orientation can be assigned. Namely, we choose one of the vertices $v_i$, $v_j$ as the positive end of $e_{ij}$ and the other one as the negative end. An orientation of a graph is called acyclic if it has no directed cycles. The degree or coordination number $k_i$ of a vertex $v_i \in V$ is the number of edges attached to it.  A $k$-regular graph is a graph with the property that each of its vertices has the same degree $k$. 

When the number of acyclic orientations $N_{AO}(G)$ grows exponentially with $v(G)$ as $v(G) \to \infty$, there exists a constant $z_G$ describing this exponential growth \cite{burton93}:
\beq
z_G = \lim_{v(G) \to \infty} \frac{\ln N_{AO}(G)}{v(G)}
\label{zdef}
\eeq
where $G$, when used as a subscript in this manner, implicitly refers to
the thermodynamic limit. We will see that the limit in Eq. (\ref{zdef}) exists for the Sierpinski gasket considered in this paper.

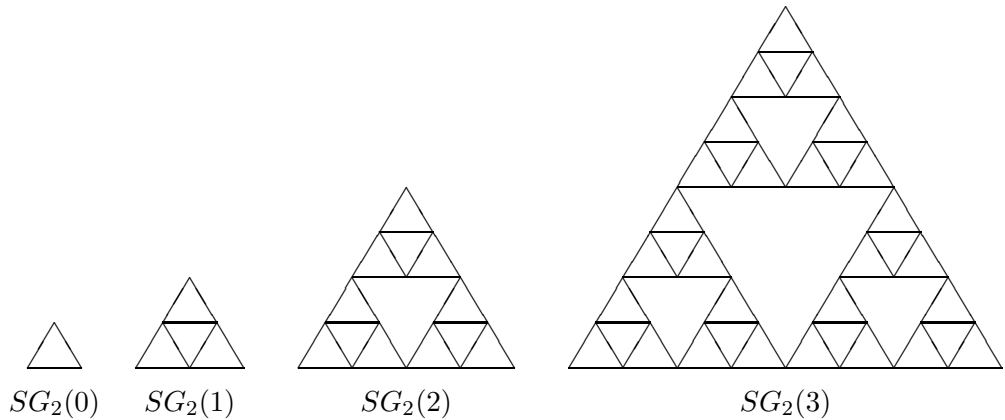
\begin{figure}[htbp]
\unitlength 1.2mm \hspace*{3mm}
\begin{picture}(108,40)
\put(0,0){\line(1,0){6}}
\put(0,0){\line(3,5){3}}
\put(6,0){\line(-3,5){3}}
\put(3,-4){\makebox(0,0){$SG_2(0)$}}
\put(12,0){\line(1,0){12}}
\put(12,0){\line(3,5){6}}
\put(24,0){\line(-3,5){6}}
\put(15,5){\line(1,0){6}}
\put(18,0){\line(3,5){3}}
\put(18,0){\line(-3,5){3}}
\put(18,-4){\makebox(0,0){$SG_2(1)$}}
\put(30,0){\line(1,0){24}}
\put(30,0){\line(3,5){12}}
\put(54,0){\line(-3,5){12}}
\put(36,10){\line(1,0){12}}
\put(42,0){\line(3,5){6}}
\put(42,0){\line(-3,5){6}}
\multiput(33,5)(12,0){2}{\line(1,0){6}}
\multiput(36,0)(12,0){2}{\line(3,5){3}}
\multiput(36,0)(12,0){2}{\line(-3,5){3}}
\put(39,15){\line(1,0){6}}
\put(42,10){\line(3,5){3}}
\put(42,10){\line(-3,5){3}}
\put(42,-4){\makebox(0,0){$SG_2(2)$}}
\put(60,0){\line(1,0){48}}
\put(72,20){\line(1,0){24}}
\put(60,0){\line(3,5){24}}
\put(84,0){\line(3,5){12}}
\put(84,0){\line(-3,5){12}}
\put(108,0){\line(-3,5){24}}
\put(66,10){\line(1,0){12}}
\put(90,10){\line(1,0){12}}
\put(78,30){\line(1,0){12}}
\put(72,0){\line(3,5){6}}
\put(96,0){\line(3,5){6}}
\put(84,20){\line(3,5){6}}
\put(72,0){\line(-3,5){6}}
\put(96,0){\line(-3,5){6}}
\put(84,20){\line(-3,5){6}}
\multiput(63,5)(12,0){4}{\line(1,0){6}}
\multiput(66,0)(12,0){4}{\line(3,5){3}}
\multiput(66,0)(12,0){4}{\line(-3,5){3}}
\multiput(69,15)(24,0){2}{\line(1,0){6}}
\multiput(72,10)(24,0){2}{\line(3,5){3}}
\multiput(72,10)(24,0){2}{\line(-3,5){3}}
\multiput(75,25)(12,0){2}{\line(1,0){6}}
\multiput(78,20)(12,0){2}{\line(3,5){3}}
\multiput(78,20)(12,0){2}{\line(-3,5){3}}
\put(81,35){\line(1,0){6}}
\put(84,30){\line(3,5){3}}
\put(84,30){\line(-3,5){3}}
\put(84,-4){\makebox(0,0){$SG_2(3)$}}
\end{picture}

\vspace*{5mm}
\caption{\footnotesize{The first four stages $n=0,1,2,3$ of the two-dimensional Sierpinski gasket $SG_2(n)$.}} 
\label{sgfig}
\end{figure}

The construction of the two-dimensional Sierpinski gasket $SG_2(n)$ at stage $n$ is shown in Fig. \ref{sgfig}. At stage $n=0$, it is an equilateral triangle; while stage $n+1$ is obtained by the juxtaposition of three $n$-stage structures. In general, the Sierpinski gaskets $SG_d$ can be built in any Euclidean dimension $d$ with fractal dimension $D=\ln(d+1)/\ln2$ \cite{Gefen81}. For the Sierpinski gasket $SG_d(n)$, the numbers of edges and vertices are given by 
\beq
e(SG_d(n)) = {d+1 \choose 2} (d+1)^n = \frac{d}{2} (d+1)^{n+1} \ ,
\label{e}
\eeq
\beq
v(SG_d(n)) = \frac{d+1}{2} [(d+1)^n+1] \ .
\label{v}
\eeq
Except the $(d+1)$ outmost vertices which have degree $d$, all other vertices of $SG_d(n)$ have degree $2d$. In the large $n$ limit, $SG_d$ is $2d$-regular. 

\bigskip

\bigskip

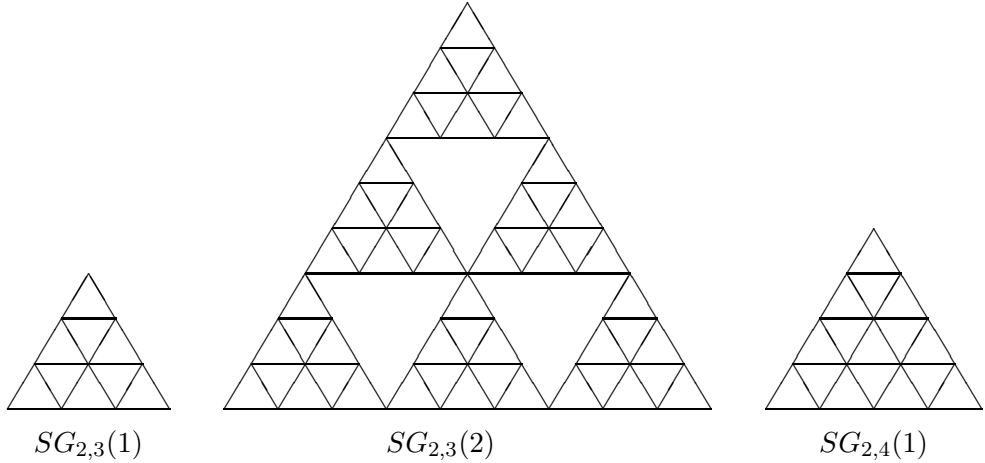
\begin{figure}[htbp]
\unitlength 1.2mm \hspace*{3mm}
\begin{picture}(108,45)
\put(0,0){\line(1,0){18}}
\put(3,5){\line(1,0){12}}
\put(6,10){\line(1,0){6}}
\put(0,0){\line(3,5){9}}
\put(6,0){\line(3,5){6}}
\put(12,0){\line(3,5){3}}
\put(18,0){\line(-3,5){9}}
\put(12,0){\line(-3,5){6}}
\put(6,0){\line(-3,5){3}}
\put(9,-4){\makebox(0,0){$SG_{2,3}(1)$}}
\put(24,0){\line(1,0){54}}
\put(33,15){\line(1,0){36}}
\put(42,30){\line(1,0){18}}
\put(24,0){\line(3,5){27}}
\put(42,0){\line(3,5){18}}
\put(60,0){\line(3,5){9}}
\put(78,0){\line(-3,5){27}}
\put(60,0){\line(-3,5){18}}
\put(42,0){\line(-3,5){9}}
\multiput(27,5)(18,0){3}{\line(1,0){12}}
\multiput(30,10)(18,0){3}{\line(1,0){6}}
\multiput(30,0)(18,0){3}{\line(3,5){6}}
\multiput(36,0)(18,0){3}{\line(3,5){3}}
\multiput(36,0)(18,0){3}{\line(-3,5){6}}
\multiput(30,0)(18,0){3}{\line(-3,5){3}}
\multiput(36,20)(18,0){2}{\line(1,0){12}}
\multiput(39,25)(18,0){2}{\line(1,0){6}}
\multiput(39,15)(18,0){2}{\line(3,5){6}}
\multiput(45,15)(18,0){2}{\line(3,5){3}}
\multiput(45,15)(18,0){2}{\line(-3,5){6}}
\multiput(39,15)(18,0){2}{\line(-3,5){3}}
\put(45,35){\line(1,0){12}}
\put(48,40){\line(1,0){6}}
\put(48,30){\line(3,5){6}}
\put(54,30){\line(3,5){3}}
\put(54,30){\line(-3,5){6}}
\put(48,30){\line(-3,5){3}}
\put(48,-4){\makebox(0,0){$SG_{2,3}(2)$}}
\put(84,0){\line(1,0){24}}
\put(87,5){\line(1,0){18}}
\put(90,10){\line(1,0){12}}
\put(93,15){\line(1,0){6}}
\put(84,0){\line(3,5){12}}
\put(90,0){\line(3,5){9}}
\put(96,0){\line(3,5){6}}
\put(102,0){\line(3,5){3}}
\put(108,0){\line(-3,5){12}}
\put(102,0){\line(-3,5){9}}
\put(96,0){\line(-3,5){6}}
\put(90,0){\line(-3,5){3}}
\put(96,-4){\makebox(0,0){$SG_{2,4}(1)$}}
\end{picture}

\vspace*{5mm}
\caption{\footnotesize{The generalized two-dimensional Sierpinski gasket $SG_{2,b}(n)$ with $b=3$ at stage $n=1, 2$ and $b=4$ at stage $n=1$.}} 
\label{sgbfig}
\end{figure}

The Sierpinski gasket can be generalized, denoted by $SG_{d,b}(n)$, by introducing the side length $b$ which is an integer larger or equal to two \cite{Hilfer}. The generalized Sierpinski gasket at stage $n+1$ is constructed from $b$ layers of stage $n$ hypertetrahedrons. The two-dimensional $SG_{2,b}(n)$ with $b=3$ at stage $n=1, 2$ and $b=4$ at stage $n=1$ are illustrated in Fig. \ref{sgbfig}. The ordinary Sierpinski gasket $SG_d(n)$ corresponds to the $b=2$ case, where the index $b$ is neglected for simplicity. The Hausdorff dimension for $SG_{d,b}$ is given by $D=\ln {b+d-1 \choose d} / \ln b$ \cite{Hilfer}. Notice that $SG_{d,b}$ is not $k$-regular even in the thermodynamic limit.

\bigskip

\bigskip

\section{The number of acyclic orientations on $SG_2(n)$}
\label{sectionIII}

In this section we derive the asymptotic growth constant for the number of acyclic orientations on the two-dimensional Sierpinski gasket $SG_2(n)$ in detail. Let us start with the definitions of the quantities to be used.

\bigskip

\begin{defi} \label{defisg2} Consider the generalized two-dimensional Sierpinski gasket $SG_{2,b}(n)$ at stage $n$. Denote the rightmost vertex as $i(n)$, the topmost vertex as $j(n)$ and the leftmost vertex as $o$. (a) Define $f_{2,b}(n) \equiv N_{AO}(SG_{2,b}(n))$ as the number of acyclic orientations. (b) Define $a_{2,b}(n)$ as the number of acyclic orientations such that there is at least one oriented path from $i(n)$ to $j(n)$, at least one oriented path from $j(n)$ to $o$ and at least one oriented path from $i(n)$ to $o$. (c) Define $b_{2,b}(n)$ as the number of acyclic orientations such that there is at least one oriented path from $i(n)$ to $j(n)$ and at least one oriented path from $i(n)$ to $o$, but there is no oriented paths between $j(n)$ and $o$. (d) Define $c_{2,b}(n)$ as the number of acyclic orientations such that there is at least one oriented path from $i(n)$ to $o$, but there is no oriented paths between $i(n)$, $j(n)$ and no oriented paths between $j(n)$, $o$. (e) Define $d_{2,b}(n)$ as the number of acyclic orientations such that there is no oriented paths between $i(n)$, $j(n)$, no oriented paths between $j(n)$, $o$ and no oriented paths between $i(n)$, $o$.
\end{defi}

\bigskip

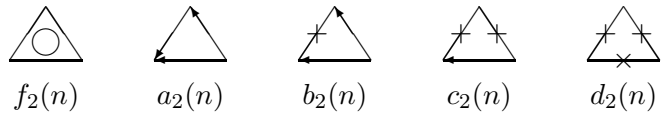
\begin{figure}[htbp]
\unitlength 2.4mm 
\begin{picture}(36,3)
\put(0,0){\line(1,0){4}}
\put(0,0){\line(2,3){2}}
\put(4,0){\line(-2,3){2}}
\put(2,1.05){\circle{1.5}}
\put(2,-2){\makebox(0,0){$f_2(n)$}}
\put(12,0){\vector(-1,0){4}}
\put(12,0){\vector(-2,3){2}}
\put(10,3){\vector(-2,-3){2}}
\put(10,-2){\makebox(0,0){$a_2(n)$}}
\put(20,0){\vector(-1,0){4}}
\put(20,0){\vector(-2,3){2}}
\put(18,3){\line(-2,-3){2}}
\put(17,1.5){\makebox(0,0){$+$}}
\put(18,-2){\makebox(0,0){$b_2(n)$}}
\put(28,0){\vector(-1,0){4}}
\put(28,0){\line(-2,3){2}}
\put(26,3){\line(-2,-3){2}}
\multiput(25,1.5)(2,0){2}{\makebox(0,0){$+$}}
\put(26,-2){\makebox(0,0){$c_2(n)$}}
\put(36,0){\line(-1,0){4}}
\put(34,3){\line(2,-3){2}}
\put(34,3){\line(-2,-3){2}}
\multiput(33,1.5)(2,0){2}{\makebox(0,0){$+$}}
\put(34,0){\makebox(0,0){$\times$}}
\put(34,-2){\makebox(0,0){$d_2(n)$}}
\end{picture}
\vspace*{5mm}

\caption{\footnotesize{Illustration for $f_2(n)$, $a_2(n)$, $b_2(n)$, $c_2(n)$, $d_2(n)$. An arrow denotes at least one oriented path from one outmost vertex to another, while a cross denotes no oriented path between the corresponding outmost vertices.}} 
\label{fabcdfig}
\end{figure}

\begin{figure}[htbp]
\unitlength 2.4mm 
\begin{picture}(44,3)
\put(4,0){\vector(-1,0){4}}
\put(4,0){\vector(-2,3){2}}
\put(2,3){\vector(-2,-3){2}}
\put(8,0){\vector(1,0){4}}
\put(8,0){\vector(2,3){2}}
\put(12,0){\vector(-2,3){2}}
\put(16,0){\vector(1,0){4}}
\put(18,3){\vector(2,-3){2}}
\put(18,3){\vector(-2,-3){2}}
\put(24,0){\vector(1,0){4}}
\put(24,0){\vector(2,3){2}}
\put(26,3){\vector(2,-3){2}}
\put(36,0){\vector(-1,0){4}}
\put(34,3){\vector(2,-3){2}}
\put(34,3){\vector(-2,-3){2}}
\put(40,0){\vector(2,3){2}}
\put(44,0){\vector(-1,0){4}}
\put(44,0){\vector(-2,3){2}}
\end{picture}

\caption{\footnotesize{Illustration for the configurations $a_2(n)$. An arrow denotes at least one oriented path from one outmost vertex to another.}} 
\label{afig}
\end{figure}
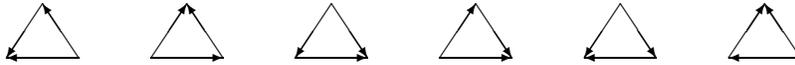

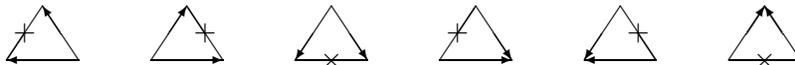
\begin{figure}[htbp]
\unitlength 2.4mm 
\begin{picture}(44,3)
\put(4,0){\vector(-1,0){4}}
\put(4,0){\vector(-2,3){2}}
\put(2,3){\line(-2,-3){2}}
\put(1,1.5){\makebox(0,0){$+$}}
\put(8,0){\vector(1,0){4}}
\put(8,0){\vector(2,3){2}}
\put(12,0){\line(-2,3){2}}
\put(11,1.5){\makebox(0,0){$+$}}
\put(16,0){\line(1,0){4}}
\put(18,3){\vector(2,-3){2}}
\put(18,3){\vector(-2,-3){2}}
\put(18,0){\makebox(0,0){$\times$}}
\put(24,0){\vector(1,0){4}}
\put(24,0){\line(2,3){2}}
\put(26,3){\vector(2,-3){2}}
\put(25,1.5){\makebox(0,0){$+$}}
\put(36,0){\vector(-1,0){4}}
\put(34,3){\line(2,-3){2}}
\put(34,3){\vector(-2,-3){2}}
\put(35,1.5){\makebox(0,0){$+$}}
\put(40,0){\vector(2,3){2}}
\put(44,0){\line(-1,0){4}}
\put(44,0){\vector(-2,3){2}}
\put(42,0){\makebox(0,0){$\times$}}
\end{picture}

\caption{\footnotesize{Illustration for the configurations $b_2(n)$. An arrow denotes at least one oriented path from one outmost vertex to another, while a cross denotes no oriented path between the corresponding outmost vertices.}} 
\label{bfig}
\end{figure}

\begin{figure}[htbp]
\unitlength 2.4mm 
\begin{picture}(44,3)
\put(4,0){\vector(-1,0){4}}
\put(4,0){\line(-2,3){2}}
\put(2,3){\line(-2,-3){2}}
\multiput(1,1.5)(2,0){2}{\makebox(0,0){$+$}}
\put(8,0){\line(1,0){4}}
\put(8,0){\vector(2,3){2}}
\put(12,0){\line(-2,3){2}}
\put(11,1.5){\makebox(0,0){$+$}}
\put(10,0){\makebox(0,0){$\times$}}
\put(16,0){\line(1,0){4}}
\put(18,3){\vector(2,-3){2}}
\put(18,3){\line(-2,-3){2}}
\put(18,0){\makebox(0,0){$\times$}}
\put(17,1.5){\makebox(0,0){$+$}}
\put(24,0){\vector(1,0){4}}
\put(24,0){\line(2,3){2}}
\put(26,3){\line(2,-3){2}}
\multiput(25,1.5)(2,0){2}{\makebox(0,0){$+$}}
\put(36,0){\line(-1,0){4}}
\put(34,3){\line(2,-3){2}}
\put(34,3){\vector(-2,-3){2}}
\put(35,1.5){\makebox(0,0){$+$}}
\put(34,0){\makebox(0,0){$\times$}}
\put(40,0){\line(2,3){2}}
\put(44,0){\line(-1,0){4}}
\put(44,0){\vector(-2,3){2}}
\put(42,0){\makebox(0,0){$\times$}}
\put(41,1.5){\makebox(0,0){$+$}}
\end{picture}

\caption{\footnotesize{Illustration for the configurations $c_2(n)$. An arrow denotes at least one oriented path from one outmost vertex to another, while a cross denotes no oriented path between the corresponding outmost vertices.}} 
\label{cfig}
\end{figure}
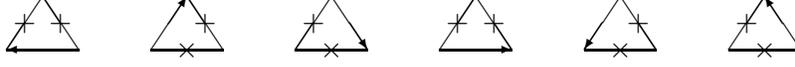

Since we only consider the ordinary Sierpinski gasket in this section, for simplicity we shall use the notations $f_2(n)$, $a_2(n)$, $b_2(n)$, $c_2(n)$ and $d_2(n)$ that are illustrated in Fig. \ref{fabcdfig}, where only the outmost vertices are shown.
As it is allowed to have all the orientations in $a_2(n)$ reversed, as well as the threefold rotation symmetry, there are six configurations that give the value $a_2(n)$ as illustrated in Fig. \ref{afig}. Similarly, there are six configurations for $b_2(n)$ and six configurations for $c_2(n)$ as illustrated in Figs. \ref{bfig} and \ref{cfig}, respectively.
It follows that
\beq
f_2(n) = 6a_2(n)+6b_2(n)+6c_2(n)+d_2(n) \ .
\label{fsg2}
\eeq
The initial values at stage zero are $a_2(0)=1$, $b_2(0)=0$, $c_2(0)=0$, $d_2(0)=0$, so that $f_2(0)=6$. The purpose of this section is to obtain the asymptotic behavior of $f_2(n)$.
The five quantities $f_2(n)$, $a_2(n)$, $b_2(n)$, $c_2(n)$ and $d_2(n)$ satisfy the recursion relations given in the following lemma. 

\bigskip

\bigskip

\bigskip

\bigskip

\bigskip

\begin{lemma} \label{lemmasg2r} For any non-negative integer $n$,
\beq
f_2(n+1) = f_2^3(n) - 2[3a_2(n)+2b_2(n)+c_2(n)]^3 \ , 
\label{feq}
\eeq
\beq
a_2(n+1) = 5a_2^3(n) + 8a_2^2(n)b_2(n) + 6a_2^2(n)c_2(n) + 3a_2(n)b_2^2(n) + 4a_2(n)b_2(n)c_2(n) + a_2(n)c_2^2(n) \ , 
\label{aeq}
\eeq
\beqs
b_2(n+1) & = & 12a_2^3(n) + 35a_2^2(n)b_2(n) + 24a_2^2(n)c_2(n) + 30a_2(n)b_2^2(n) + 34a_2(n)b_2(n)c_2(n) \cr\cr
& & + 4a_2(n)c_2^2(n) + 8b_2^3(n) + 12b_2^2(n)c_2(n) + 3b_2(n)c_2^2(n) \ , 
\label{beq}
\eeqs
\beqs
c_2(n+1) & = & 6a_2^3(n) + 26a_2^2(n)b_2(n) + 39a_2^2(n)c_2(n) + 9a_2^2(n)d_2(n) + 30a_2(n)b_2^2(n) \cr\cr
& & + 76a_2(n)b_2(n)c_2(n) + 12a_2(n)b_2(n)d_2(n) + 40a_2(n)c_2^2(n) + 6a_2(n)c_2(n)d_2(n) \cr\cr
& & + 10b_2^3(n) + 33b_2^2(n)c_2(n) + 4b_2^2(n)d_2(n) + 30b_2(n)c_2^2(n) + 4b_2(n)c_2(n)d_2(n) \cr\cr
& & + 7c_2^3(n) + c_2^2(n)d_2(n) \ ,
\label{ceq}
\eeqs
\beqs
d_2(n+1) & = & 24a_2^3(n) + 126a_2^2(n)b_2(n) + 180a_2^2(n)c_2(n) + 54a_2^2(n)d_2(n) + 198a_2(n)b_2^2(n) \cr\cr
& & + 540a_2(n)b_2(n)c_2(n) + 144a_2(n)b_2(n)d_2(n) + 360a_2(n)c_2^2(n) \cr\cr
& & + 180a_2(n)c_2(n)d_2(n) + 18a_2(n)d_2^2(n) + 92b_2^3(n) + 354b_2^2(n)c_2(n) \cr\cr
& & + 84b_2^2(n)d_2(n) + 438b_2(n)c_2^2(n) + 192b_2(n)c_2(n)d_2(n) + 18b_2(n)d_2^2(n) \cr\cr
& & + 172c_2^3(n) + 102c_2^2(n)d_2(n) + 18c_2(n)d_2^2(n) + d_2^3(n) \ .
\label{deq}
\eeqs
\end{lemma}

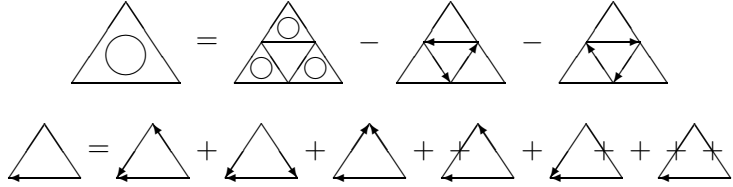
\begin{figure}[htbp]
\unitlength 1.8mm 
\begin{picture}(44,6)
\put(0,0){\line(1,0){8}}
\put(0,0){\line(2,3){4}}
\put(8,0){\line(-2,3){4}}
\put(4,2.1){\circle{3}}
\put(10,3){\makebox(0,0){$=$}}
\put(12,0){\line(1,0){8}}
\put(12,0){\line(2,3){4}}
\put(16,0){\line(2,3){2}}
\put(16,0){\line(-2,3){2}}
\put(14,3){\line(1,0){4}}
\put(20,0){\line(-2,3){4}}
\multiput(14,1.1)(4,0){2}{\circle{1.5}}
\put(16,4.1){\circle{1.5}}
\put(22,3){\makebox(0,0){$-$}}
\put(24,0){\line(2,3){4}}
\put(24,0){\line(1,0){8}}
\put(28,0){\vector(2,3){2}}
\put(26,3){\vector(2,-3){2}}
\put(30,3){\vector(-1,0){4}}
\put(32,0){\line(-2,3){4}}
\put(34,3){\makebox(0,0){$-$}}
\put(36,0){\line(2,3){4}}
\put(36,0){\line(1,0){8}}
\put(42,3){\vector(-2,-3){2}}
\put(40,0){\vector(-2,3){2}}
\put(38,3){\vector(1,0){4}}
\put(44,0){\line(-2,3){4}}
\end{picture}

\vspace*{5mm}
\unitlength 2.4mm 
\begin{picture}(40,3)
\put(4,0){\vector(-1,0){4}}
\put(0,0){\line(2,3){2}}
\put(4,0){\line(-2,3){2}}
\put(5,1.5){\makebox(0,0){$=$}}
\put(10,0){\vector(-1,0){4}}
\put(8,3){\vector(-2,-3){2}}
\put(10,0){\vector(-2,3){2}}
\put(11,1.5){\makebox(0,0){$+$}}
\put(16,0){\vector(-1,0){4}}
\put(14,3){\vector(2,-3){2}}
\put(14,3){\vector(-2,-3){2}}
\put(17,1.5){\makebox(0,0){$+$}}
\put(22,0){\vector(-1,0){4}}
\put(18,0){\vector(2,3){2}}
\put(22,0){\vector(-2,3){2}}
\put(23,1.5){\makebox(0,0){$+$}}
\put(28,0){\vector(-1,0){4}}
\put(24,0){\line(2,3){2}}
\put(28,0){\vector(-2,3){2}}
\put(25,1.5){\makebox(0,0){$+$}}
\put(29,1.5){\makebox(0,0){$+$}}
\put(34,0){\vector(-1,0){4}}
\put(32,3){\vector(-2,-3){2}}
\put(34,0){\line(-2,3){2}}
\put(33,1.5){\makebox(0,0){$+$}}
\put(35,1.5){\makebox(0,0){$+$}}
\put(40,0){\vector(-1,0){4}}
\put(36,0){\line(2,3){2}}
\put(40,0){\line(-2,3){2}}
\multiput(37,1.5)(2,0){2}{\makebox(0,0){$+$}}
\end{picture}

\caption{\footnotesize{Illustrations for the expression of $f_2(n+1)$, and the representation of a solid line.}} 
\label{ffig}
\end{figure}

{\sl Proof} \quad 
Let us establish Eq. (\ref{feq}) first. The Sierpinski gasket $SG_2(n+1)$ is composed of three $SG_2(n)$ with three pairs of vertices identified. For the number $f_2(n+1)$, the unallowable configurations are those with a directed cycle around the lacunary triangle as illustrated in Fig. \ref{ffig}. For each of the constituting $SG_2(n)$, if one only knows that at least one oriented path from $i(n)$ to $o$ without specifying the relation between $i(n)$, $j(n)$ and that between $j(n)$, $o$, the number is given as $3a_2(n)+2b_2(n)+c_2(n)$ that is also shown in Fig. \ref{ffig} to verify Eq. (\ref{feq}). 

\bigskip

\bigskip

\begin{figure}[htbp]
\unitlength 1.8mm 
\begin{picture}(80,6)
\put(8,0){\vector(-1,0){8}}
\put(8,0){\vector(-2,3){4}}
\put(4,6){\vector(-2,-3){4}}
\put(10,3){\makebox(0,0){$=$}}
\put(16,6){\vector(-2,-3){2}}
\put(18,3){\vector(-2,3){2}}
\put(18,3){\vector(-1,0){4}}
\put(14,3){\vector(-2,-3){2}}
\put(16,0){\vector(-2,3){2}}
\put(16,0){\vector(-1,0){4}}
\put(16,0){\vector(2,3){2}}
\put(16,0){\vector(1,0){4}}
\put(20,0){\vector(-2,3){2}}
\put(22,3){\makebox(0,0){$+$}}
\put(28,6){\vector(-2,-3){2}}
\put(30,3){\vector(-2,3){2}}
\put(30,3){\vector(-1,0){4}}
\put(26,3){\vector(-2,-3){2}}
\put(28,0){\line(-2,3){2}}
\put(27,1.5){\makebox(0,0){$+$}}
\put(28,0){\vector(-1,0){4}}
\put(28,0){\vector(2,3){2}}
\put(28,0){\vector(1,0){4}}
\put(32,0){\vector(-2,3){2}}
\put(34,3){\makebox(0,0){$+$}}
\put(40,6){\vector(-2,-3){2}}
\put(42,3){\vector(-2,3){2}}
\put(42,3){\vector(-1,0){4}}
\put(38,3){\vector(-2,-3){2}}
\put(38,3){\vector(2,-3){2}}
\put(36,0){\vector(1,0){4}}
\put(42,3){\vector(-2,-3){2}}
\put(44,0){\vector(-1,0){4}}
\put(44,0){\vector(-2,3){2}}
\put(46,3){\makebox(0,0){$+$}}
\put(52,6){\vector(-2,-3){2}}
\put(54,3){\vector(-2,3){2}}
\put(54,3){\vector(-1,0){4}}
\put(50,3){\vector(-2,-3){2}}
\put(50,3){\vector(2,-3){2}}
\put(48,0){\vector(1,0){4}}
\put(54,3){\line(-2,-3){2}}
\put(53,1.5){\makebox(0,0){$+$}}
\put(56,0){\vector(-1,0){4}}
\put(56,0){\vector(-2,3){2}}
\put(58,3){\makebox(0,0){$+$}}
\put(64,6){\vector(-2,-3){2}}
\put(66,3){\vector(-2,3){2}}
\put(66,3){\vector(-1,0){4}}
\put(62,3){\vector(-2,-3){2}}
\put(64,0){\vector(-2,3){2}}
\put(64,0){\vector(-1,0){4}}
\put(66,3){\vector(-2,-3){2}}
\put(68,0){\vector(-1,0){4}}
\put(68,0){\vector(-2,3){2}}
\put(70,3){\makebox(0,0){$+$}}
\put(76,6){\vector(-2,-3){2}}
\put(78,3){\vector(-2,3){2}}
\put(78,3){\vector(-1,0){4}}
\put(74,3){\vector(-2,-3){2}}
\put(76,0){\vector(-2,3){2}}
\put(76,0){\vector(-1,0){4}}
\put(76,0){\vector(2,3){2}}
\put(80,0){\vector(-1,0){4}}
\put(80,0){\vector(-2,3){2}}
\end{picture}

\vspace*{2mm}
\begin{picture}(80,6)
\put(10,3){\makebox(0,0){$+$}}
\put(16,6){\vector(-2,-3){2}}
\put(18,3){\vector(-2,3){2}}
\put(18,3){\vector(-1,0){4}}
\put(14,3){\vector(-2,-3){2}}
\put(16,0){\vector(-2,3){2}}
\put(16,0){\vector(-1,0){4}}
\put(16,0){\line(2,3){2}}
\put(17,1.5){\makebox(0,0){$+$}}
\put(20,0){\vector(-1,0){4}}
\put(20,0){\vector(-2,3){2}}
\put(22,3){\makebox(0,0){$+$}}
\put(28,6){\vector(-2,-3){2}}
\put(30,3){\vector(-2,3){2}}
\put(30,3){\vector(-1,0){4}}
\put(26,3){\vector(-2,-3){2}}
\put(26,3){\vector(2,-3){2}}
\put(28,0){\vector(-1,0){4}}
\put(30,3){\vector(-2,-3){2}}
\put(32,0){\vector(-1,0){4}}
\put(32,0){\vector(-2,3){2}}
\put(34,3){\makebox(0,0){$+$}}
\put(40,6){\vector(-2,-3){2}}
\put(42,3){\vector(-2,3){2}}
\put(42,3){\vector(-1,0){4}}
\put(38,3){\vector(-2,-3){2}}
\put(38,3){\vector(2,-3){2}}
\put(40,0){\vector(-1,0){4}}
\put(42,3){\line(-2,-3){2}}
\put(41,1.5){\makebox(0,0){$+$}}
\put(44,0){\vector(-1,0){4}}
\put(44,0){\vector(-2,3){2}}
\put(46,3){\makebox(0,0){$+$}}
\put(52,6){\vector(-2,-3){2}}
\put(54,3){\vector(-2,3){2}}
\put(54,3){\vector(-1,0){4}}
\put(50,3){\vector(-2,-3){2}}
\put(52,0){\line(-2,3){2}}
\put(51,1.5){\makebox(0,0){$+$}}
\put(52,0){\vector(-1,0){4}}
\put(54,3){\vector(-2,-3){2}}
\put(56,0){\vector(-1,0){4}}
\put(56,0){\vector(-2,3){2}}
\put(58,3){\makebox(0,0){$+$}}
\put(64,6){\vector(-2,-3){2}}
\put(66,3){\vector(-2,3){2}}
\put(66,3){\vector(-1,0){4}}
\put(62,3){\vector(-2,-3){2}}
\put(64,0){\line(-2,3){2}}
\put(63,1.5){\makebox(0,0){$+$}}
\put(64,0){\vector(-1,0){4}}
\put(64,0){\vector(2,3){2}}
\put(68,0){\vector(-1,0){4}}
\put(68,0){\vector(-2,3){2}}
\put(70,3){\makebox(0,0){$+$}}
\put(76,6){\vector(-2,-3){2}}
\put(78,3){\vector(-2,3){2}}
\put(78,3){\vector(-1,0){4}}
\put(74,3){\vector(-2,-3){2}}
\put(74,3){\line(2,-3){2}}
\put(75,1.5){\makebox(0,0){$+$}}
\put(76,0){\vector(-1,0){4}}
\put(78,3){\line(-2,-3){2}}
\put(77,1.5){\makebox(0,0){$+$}}
\put(80,0){\vector(-1,0){4}}
\put(80,0){\vector(-2,3){2}}
\end{picture}

\vspace*{2mm}
\begin{picture}(80,6)
\put(10,3){\makebox(0,0){$+$}}
\put(16,6){\vector(-2,-3){2}}
\put(18,3){\vector(-2,3){2}}
\put(18,3){\vector(-1,0){4}}
\put(14,3){\vector(-2,-3){2}}
\put(16,0){\vector(-2,3){2}}
\put(16,0){\vector(-1,0){4}}
\put(16,0){\vector(2,3){2}}
\put(20,0){\vector(-2,3){2}}
\put(16,0){\line(1,0){4}}
\put(18,0){\makebox(0,0){$\times$}}
\put(22,3){\makebox(0,0){$+$}}
\put(28,6){\vector(-2,-3){2}}
\put(30,3){\vector(-2,3){2}}
\put(30,3){\vector(-1,0){4}}
\put(26,3){\vector(-2,-3){2}}
\put(26,3){\vector(2,-3){2}}
\put(24,0){\line(1,0){4}}
\put(26,0){\makebox(0,0){$\times$}}
\put(30,3){\vector(-2,-3){2}}
\put(32,0){\vector(-1,0){4}}
\put(32,0){\vector(-2,3){2}}
\put(34,3){\makebox(0,0){$+$}}
\put(40,6){\vector(-2,-3){2}}
\put(42,3){\vector(-2,3){2}}
\put(42,3){\vector(-1,0){4}}
\put(38,3){\vector(-2,-3){2}}
\put(40,0){\line(-2,3){2}}
\put(39,1.5){\makebox(0,0){$+$}}
\put(40,0){\vector(-1,0){4}}
\put(40,0){\vector(2,3){2}}
\put(44,0){\line(-1,0){4}}
\put(42,0){\makebox(0,0){$\times$}}
\put(44,0){\vector(-2,3){2}}
\put(46,3){\makebox(0,0){$+$}}
\put(52,6){\vector(-2,-3){2}}
\put(54,3){\vector(-2,3){2}}
\put(54,3){\vector(-1,0){4}}
\put(50,3){\vector(-2,-3){2}}
\put(50,3){\vector(2,-3){2}}
\put(52,0){\line(-1,0){4}}
\put(50,0){\makebox(0,0){$\times$}}
\put(52,0){\line(2,3){2}}
\put(53,1.5){\makebox(0,0){$+$}}
\put(56,0){\vector(-1,0){4}}
\put(56,0){\vector(-2,3){2}}
\put(58,3){\makebox(0,0){$+$}}
\put(64,6){\vector(-2,-3){2}}
\put(66,3){\vector(-2,3){2}}
\put(66,3){\vector(-1,0){4}}
\put(62,3){\vector(-2,-3){2}}
\put(64,0){\vector(-2,3){2}}
\put(64,0){\vector(-1,0){4}}
\put(66,3){\line(-2,-3){2}}
\put(65,1.5){\makebox(0,0){$+$}}
\put(68,0){\line(-1,0){4}}
\put(66,0){\makebox(0,0){$\times$}}
\put(68,0){\vector(-2,3){2}}
\put(70,3){\makebox(0,0){$+$}}
\put(76,6){\vector(-2,-3){2}}
\put(78,3){\vector(-2,3){2}}
\put(78,3){\vector(-1,0){4}}
\put(74,3){\vector(-2,-3){2}}
\put(74,3){\line(2,-3){2}}
\put(75,1.5){\makebox(0,0){$+$}}
\put(76,0){\line(-1,0){4}}
\put(74,0){\makebox(0,0){$\times$}}
\put(78,3){\vector(-2,-3){2}}
\put(80,0){\vector(-1,0){4}}
\put(80,0){\vector(-2,3){2}}
\end{picture}

\vspace*{2mm}
\begin{picture}(80,6)
\put(10,3){\makebox(0,0){$+$}}
\put(16,6){\vector(-2,-3){2}}
\put(18,3){\vector(-2,3){2}}
\put(18,3){\vector(-1,0){4}}
\put(14,3){\vector(-2,-3){2}}
\put(14,3){\vector(2,-3){2}}
\put(16,0){\vector(-1,0){4}}
\put(16,0){\line(2,3){2}}
\put(17,1.5){\makebox(0,0){$+$}}
\put(20,0){\vector(-2,3){2}}
\put(16,0){\line(1,0){4}}
\put(18,0){\makebox(0,0){$\times$}}
\put(22,3){\makebox(0,0){$+$}}
\put(28,6){\vector(-2,-3){2}}
\put(30,3){\vector(-2,3){2}}
\put(30,3){\vector(-1,0){4}}
\put(26,3){\vector(-2,-3){2}}
\put(26,3){\vector(2,-3){2}}
\put(24,0){\vector(1,0){4}}
\put(30,3){\line(-2,-3){2}}
\put(29,1.5){\makebox(0,0){$+$}}
\put(32,0){\line(-1,0){4}}
\put(30,0){\makebox(0,0){$\times$}}
\put(32,0){\vector(-2,3){2}}
\put(34,3){\makebox(0,0){$+$}}
\put(40,6){\vector(-2,-3){2}}
\put(42,3){\vector(-2,3){2}}
\put(42,3){\vector(-1,0){4}}
\put(38,3){\vector(-2,-3){2}}
\put(38,3){\vector(2,-3){2}}
\put(40,0){\line(-1,0){4}}
\put(38,0){\makebox(0,0){$\times$}}
\put(40,0){\line(2,3){2}}
\put(41,1.5){\makebox(0,0){$+$}}
\put(44,0){\line(-1,0){4}}
\put(42,0){\makebox(0,0){$\times$}}
\put(44,0){\vector(-2,3){2}}
\put(46,3){\makebox(0,0){$+$}}
\put(52,6){\vector(-2,-3){2}}
\put(54,3){\vector(-2,3){2}}
\put(54,3){\vector(-1,0){4}}
\put(50,3){\vector(-2,-3){2}}
\put(50,3){\line(2,-3){2}}
\put(51,1.5){\makebox(0,0){$+$}}
\put(52,0){\line(-1,0){4}}
\put(50,0){\makebox(0,0){$\times$}}
\put(52,0){\vector(2,3){2}}
\put(56,0){\vector(-1,0){4}}
\put(56,0){\vector(-2,3){2}}
\put(58,3){\makebox(0,0){$+$}}
\put(64,6){\vector(-2,-3){2}}
\put(66,3){\vector(-2,3){2}}
\put(66,3){\vector(-1,0){4}}
\put(62,3){\vector(-2,-3){2}}
\put(64,0){\line(-2,3){2}}
\put(63,1.5){\makebox(0,0){$+$}}
\put(64,0){\line(-1,0){4}}
\put(62,0){\makebox(0,0){$\times$}}
\put(64,0){\vector(2,3){2}}
\put(64,0){\vector(1,0){4}}
\put(68,0){\vector(-2,3){2}}
\put(70,3){\makebox(0,0){$+$}}
\put(76,6){\vector(-2,-3){2}}
\put(78,3){\vector(-2,3){2}}
\put(78,3){\vector(-1,0){4}}
\put(74,3){\vector(-2,-3){2}}
\put(74,3){\line(2,-3){2}}
\put(75,1.5){\makebox(0,0){$+$}}
\put(76,0){\line(-1,0){4}}
\put(74,0){\makebox(0,0){$\times$}}
\put(76,0){\vector(2,3){2}}
\put(80,0){\line(-1,0){4}}
\put(78,0){\makebox(0,0){$\times$}}
\put(80,0){\vector(-2,3){2}}
\end{picture}

\vspace*{2mm}
\begin{picture}(80,6)
\put(10,3){\makebox(0,0){$+$}}
\put(16,6){\vector(-2,-3){2}}
\put(18,3){\vector(-2,3){2}}
\put(18,3){\vector(-1,0){4}}
\put(14,3){\vector(-2,-3){2}}
\put(14,3){\line(2,-3){2}}
\put(15,1.5){\makebox(0,0){$+$}}
\put(16,0){\vector(-1,0){4}}
\put(16,0){\line(2,3){2}}
\put(17,1.5){\makebox(0,0){$+$}}
\put(20,0){\vector(-2,3){2}}
\put(16,0){\line(1,0){4}}
\put(18,0){\makebox(0,0){$\times$}}
\put(22,3){\makebox(0,0){$+$}}
\put(28,6){\vector(-2,-3){2}}
\put(30,3){\vector(-2,3){2}}
\put(30,3){\vector(-1,0){4}}
\put(26,3){\vector(-2,-3){2}}
\put(26,3){\line(2,-3){2}}
\put(27,1.5){\makebox(0,0){$+$}}
\put(24,0){\line(1,0){4}}
\put(26,0){\makebox(0,0){$\times$}}
\put(30,3){\line(-2,-3){2}}
\put(29,1.5){\makebox(0,0){$+$}}
\put(32,0){\vector(-1,0){4}}
\put(32,0){\vector(-2,3){2}}
\put(34,3){\makebox(0,0){$+$}}
\put(40,6){\vector(-2,-3){2}}
\put(42,3){\vector(-2,3){2}}
\put(42,3){\vector(-1,0){4}}
\put(38,3){\vector(-2,-3){2}}
\put(38,3){\line(2,-3){2}}
\put(39,1.5){\makebox(0,0){$+$}}
\put(40,0){\line(-1,0){4}}
\put(38,0){\makebox(0,0){$\times$}}
\put(40,0){\line(2,3){2}}
\put(41,1.5){\makebox(0,0){$+$}}
\put(44,0){\line(-1,0){4}}
\put(42,0){\makebox(0,0){$\times$}}
\put(44,0){\vector(-2,3){2}}
\end{picture}

\caption{\footnotesize{Illustration for the expression of $a_2(n+1)$.}} 
\label{aeqfig}
\end{figure}
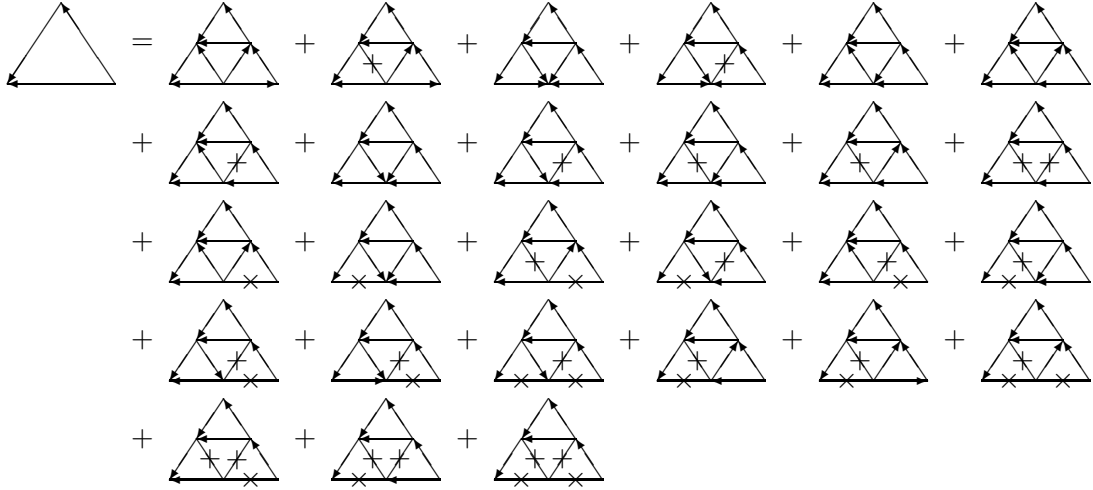

The number $a_2(n+1)$ consists of (i) five configurations where all three constituting $SG_2(n)$'s belong to the class that is enumerated by $a_2(n)$; (ii) eight configurations where two of the $SG_2(n)$'s belong to the class enumerated by $a_2(n)$ and the other one belongs to the class enumerated by $b_2(n)$; (iii) six configurations where two of the $SG_2(n)$'s belong to the class enumerated by $a_2(n)$ and the other one belongs to the class enumerated by $c_2(n)$; (iv) three configurations where one of the $SG_2(n)$'s belongs to the class enumerated by $a_2(n)$ and the other two belong to the class enumerated by $b_2(n)$; (v) four configurations where one of the $SG_2(n)$'s belongs to the class enumerated by $a_2(n)$, another one belongs to the class enumerated by $b_2(n)$ and the other one belongs to the class enumerated by $c_2(n)$; (vi) one configuration where one of the $SG_2(n)$'s belongs to the class enumerated by $a_2(n)$ and the other two belong to the class enumerated by $c_2(n)$. All these possibilities are illustrated in Fig. \ref{aeqfig}, such that Eq. (\ref{aeq}) is verified.

\bigskip

\bigskip

Similarly, the expressions of $b_2(n+1)$, $c_2(n+1)$, $d_2(n+1)$ for $SG_2(n+1)$ can be obtained with appropriate configurations of its three constituting $SG_2(n)$'s in order to verify Eqs. (\ref{beq}) - (\ref{deq}).
Eq. (\ref{feq}) can also be obtained by substituting Eqs. (\ref{aeq}) - (\ref{deq}) into Eq. (\ref{fsg2}). 
\ $\Box$

\bigskip

The values of $f_2(n)$, $a_2(n)$, $b_2(n)$, $c_2(n)$, $d_2(n)$ for small $n$ can be evaluated recursively by Eqs. (\ref{feq}) - (\ref{deq}) as listed in Table \ref{tablesg2}. These numbers grow exponentially, and do not have simple integer factorizations. To estimate the value of the asymptotic growth constant defined in Eq. (\ref{zdef}), we need the following lemma.

\bigskip

\begin{table}[htbp]
\caption{\label{tablesg2} The first few values of $f_2(n)$, $a_2(n)$, $b_2(n)$, $c_2(n)$, $d_2(n)$.}
\begin{center}
\begin{tabular}{|c||r|r|r|r|}
\hline\hline 
$n$      & 0 &   1 &         2 &                          3 \\ \hline\hline 
$f_2(n)$ & 6 & 162 & 4,069,278 & 67,294,670,068,124,357,202 \\ \hline 
$a_2(n)$ & 1 &   5 &     7,705 &            900773426769005 \\ \hline 
$b_2(n)$ & 0 &  12 &    75,648 &     29,379,570,130,675,692 \\ \hline 
$c_2(n)$ & 0 &   6 &   179,424 &    483,507,248,802,250,206 \\ \hline
$d_2(n)$ & 0 &  24 & 2,492,616 & 64,211,944,513,966,187,784 \\ \hline\hline 
\end{tabular}
\end{center}
\end{table}

\bigskip

\begin{lemma} \label{lemmasg2b} The asymptotic growth constant for the number of acyclic orientations on $SG_2(n)$ is bounded:
\beq
\frac{2}{3^{m+1}} \ln d_2(m) < z_{SG_2} < \frac{2}{3^{m+1}} \ln f_2(m) \ ,
\label{zsg2}
\eeq
where $m$ is a positive integer. The upper and lower bounds are close to each other when $m$ is large.
\end{lemma}

{\sl Proof} \quad 
We first show that the ratio $a_2(n)/b_2(n)$ is a strictly decreasing sequence for positive $n$. By Eqs. (\ref{aeq}) and (\ref{beq}), we have
\beqs
\lefteqn{ \frac{a_2(n+1)}{b_2(n+1)}} \cr\cr
& < & \frac {a_2(n) [ 5a_2^2(n) + 8a_2(n)b_2(n) + 6a_2(n)c_2(n) + 3b_2^2(n) + 4b_2(n)c_2(n) + c_2^2(n) ]} {b_2(n) [ 35a_2^2(n) + 30a_2(n)b_2(n) + 34a_2(n)c_2(n) + 8b_2^2(n) + 12b_2(n)c_2(n) + 3c_2^2(n) ]} \cr\cr
& < & \frac{3a_2(n)}{8b_2(n)} \ .
\eeqs
From the values in Table \ref{tablesg2}, $a_2(n)/b_2(n)$ is less than one for $n>0$, and it is clear that this ratio approaches to zero as $n$ increases. Similarly, $b_2(n)/c_2(n)$ is also a strictly decreasing sequence by Eqs. (\ref{beq}) and (\ref{ceq}).
\beqs
\lefteqn{ \frac{b_2(n+1)}{c_2(n+1)} } \cr\cr
& < & \{ a_2(n) [12a_2^2(n) + 24a_2(n)c_2(n) + 4c_2^2(n)] + b_2(n) [35a_2^2(n) + 30a_2(n)b_2(n) \cr\cr
& & + 34a_2(n)c_2(n) + 8b_2^2(n) + 12b_2(n)c_2(n) + 3c_2^2(n)] \} / \{ d_2(n) [6a_2(n)c_2(n) + c_2^2(n)] \cr\cr
& & + c_2(n) [39a_2^2(n) + 76a_2(n)b_2(n) + 40a_2(n)c_2(n) + 33b_2^2(n) + 30b_2(n)c_2(n) + 7c_2^2(n) ] \} \cr\cr
& < & \frac {b_2(n) [35a_2^2(n) + 42a_2(n)b_2(n) + 42a_2(n)c_2(n) + 8b_2^2(n) + 28b_2(n)c_2(n) + 7c_2^2(n)]} {c_2(n) [39a_2^2(n) + 76a_2(n)b_2(n) + 46a_2(n)c_2(n) + 33b_2^2(n) + 30b_2(n)c_2(n) + 8c_2^2(n)]} \cr\cr
& < & \frac{14b_2(n)}{15c_2(n)} \ , 
\eeqs
where we have repeatedly used the fact that $a_2(n) < b_2(n)$ given above, as well as $c_2(n) < d_2(n)$ below using Eqs. (\ref{beq}) and (\ref{ceq}).
\beqs
\lefteqn{ \frac{c_2(n+1)}{d_2(n+1)} } \cr\cr
& < & \{ 6a_2^3(n) + 26a_2^2(n)b_2(n) + 9a_2^2(n)d_2(n) + 30a_2(n)b_2^2(n) + 12a_2(n)b_2(n)d_2(n) + 10b_2^3(n) \cr\cr
& & + 4b_2^2(n)d_2(n) + c_2(n) [39a_2^2(n) + 76a_2(n)b_2(n) + 40a_2(n)c_2(n) + 6a_2(n)d_2(n) + 33b_2^2(n) \cr\cr
& & + 30b_2(n)c_2(n) + 4b_2(n)d_2(n) + 7c_2^2(n) + c_2(n)d_2(n) ] \} / \{ d_2(n) [54a_2^2(n) + 144a_2(n)b_2(n) \cr\cr
& & + 180a_2(n)c_2(n) + 18a_2(n)d_2(n) + 84b_2^2(n) + 192b_2(n)c_2(n) + 18b_2(n)d_2(n) + 102c_2^2(n) \cr\cr
& & + 18c_2(n)d_2(n) ] \} \cr\cr
& < & \{ c_2(n) [45a_2^2(n) + 102a_2(n)b_2(n) + 40a_2(n)c_2(n) + 15a_2(n)d_2(n) + 43b_2^2(n) \cr\cr
& & + 60b_2(n)c_2(n) + 8b_2(n)d_2(n) + 7c_2^2(n) + 13c_2(n)d_2(n) ] \} / \cr\cr
& & \{ d_2(n) [54a_2^2(n) + 144a_2(n)b_2(n) + 180a_2(n)c_2(n) + 18a_2(n)d_2(n) + 84b_2^2(n) \cr\cr
& & + 192b_2(n)c_2(n) + 18b_2(n)d_2(n) + 102c_2^2(n) + 18c_2(n)d_2(n) ] \} \cr\cr
& < & \frac{5c_2(n)}{6d_2(n)} \qquad \rm{for} \ n>1 \ ,
\eeqs
where we have used $a_2(n) < b_2(n) < c_2(n)$ for $n>1$. Both $b_2(n)/c_2(n)$ and $c_2(n)/d_2(n)$ approach to zero as $n$ increases. The relation $a_2(n) \ll b_2(n) \ll c_2(n) \ll d_2(n)$ for large $n$ is expectable since it is rare to keep the oriented path from one outmost vertex to another and $d_2(n)$ should dominate when $n$ becomes large. In fact, $a_2(n)$, $b_2(n)$ and $c_2(n)$ are negligible compared with $d_2(n)$ for large $n$ such that $f_2(n) \sim d_2(n)$ in the large $n$ limit. By Eqs. (\ref{feq}) and (\ref{deq}), we have the upper and lower bounds for $f_2(n)$:
\beq
d_2^3(n-1) < d_2(n) < f_2(n) < f_2^3(n-1) \ ,
\eeq
such that
\beq
d_2(m)^{3^{n-m}} < f_2(n) < f_2(m)^{3^{n-m}} \ ,
\eeq
where $m$ is a fixed integer smaller than $n$. With the definition for $z_{SG_2}$ given in Eq. (\ref{zdef}) and the number of vertices of $SG_2(n)$ is $3(3^n+1)/2$ by Eq. (\ref{v}), the proof is completed.
\ $\Box$

\bigskip

\begin{propo} \label{proposg2} The asymptotic growth constant for the number of acyclic orientations on the two-dimensional Sierpinski gasket $SG_2(n)$ in the large $n$ limit is $z_{SG_2}=1.127299070536616...$.

\end{propo}

{\sl Proof} \quad 
Define ratios $\alpha (n) \equiv a_2(n)/f_2(n)$, $\beta (n) \equiv b_2(n)/f_2(n)$ and $\gamma (n) \equiv c_2(n)/f_2(n)$. By Eq. (\ref{fsg2}), it is clear that $0 \le \alpha (n) + \beta (n) + \gamma (n) < 1$. As seen in the proof of Lemma \ref{lemmasg2b}, $\alpha (n) + \beta (n) + \gamma (n)$ is a strictly decreasing sequence. By Eq. (\ref{feq}), let us define 
\beq
r(n) \equiv \frac{f_2(n)}{f_2^3(n-1)} = 1 - 2[3\alpha (n-1) + 2\beta (n-1) + \gamma (n-1)]^3
\eeq
for positive integer $n$ . It follows that
\beqs
\ln f_2(n) & = & 3\ln f_2(n-1) + \ln r(n) = ... \cr\cr
& = & 3^{n-m} \ln f_2(m) + \sum_{j=m+1}^n 3^{n-j} \ln r(j) \cr\cr
& > & 3^{n-m} \ln f_2(m) + \left ( \frac{3^{n-m}-1}{2} \right ) \ln r(m+1) \ .
\eeqs
Divide this equation by $3(3^n+1)/2$ and take the limit $n \to \infty$, the difference between the upper bound in Eq. (\ref{zsg2}) and the asymptotic growth constant is bounded:
\beq
\frac{2}{3^{m+1}} \ln f_2(m) - z_{SG_2} \le \frac{-1}{3^{m+1}} \ln \left ( 1 - 2[3\alpha (m) + 2\beta (m) + \gamma (m)]^3 \right ) \ .
\label{zsg2d}
\eeq
When $m$ is as small as three, the right-hand-side of Eq. (\ref{zsg2d}) is about $10^{-8}$ by the values given in Table \ref{tablesg2}. Similarly, it can be shown that the difference between $z_{SG_2}$ and the lower bound (left-hand-side of Eq. (\ref{zsg2})) quickly converges to zero as $m$ increases. In another word, the numerical values of $\ln f_2(m)$ and $\ln d_2(m)$ are almost the same except for the first few $m$, and the upper and lower bounds in Eq. (\ref{zsg2}) converge to the quoted value of $z_{SG_2}$. In fact, one obtains the numerical value of $z_{SG_2}$ with more than a hundred significant figures accurate when $m$ is equal to nine. The rate of convergence will be discussed further in Section \ref{upperbounds}.
\ $\Box$

\bigskip

\section{The number of acyclic orientations on $SG_{2,b}(n)$ with $b=3$} 
\label{sectionIV}

The method given in the previous section can be applied to the number of acyclic orientations on $SG_{d,b}(n)$ with larger values of $d$ and $b$. The number of configurations to be considered increases as $d$ and $b$ increase, and the recursion relations must be derived individually for each $d$ and $b$. 
In this section, we consider the generalized two-dimensional Sierpinski gasket $SG_{2,b}(n)$ with the number of layers $b$ equal to three. 
For $SG_{2,3}(n)$, the numbers of edges and vertices are given by 
\beq
e(SG_{2,3}(n)) = 3 \times 6^n \ ,
\label{esg23}
\eeq
\beq
v(SG_{2,3}(n)) = \frac{7 \times 6^n + 8}{5} \ ,
\label{vsg23}
\eeq
where the three outmost vertices have degree two. There are $(6^n-1)/5$ vertices of $SG_{2,3}(n)$ with degree six and $6(6^n-1)/5$ vertices with degree four. By Definition \ref{defisg2}, the number of acyclic orientations is $f_{2,3}(n) = 6[a_{2,3}(n)+b_{2,3}(n)+c_{2,3}(n)]+d_{2,3}(n)$. The initial values are the same as for $SG_2$: $a_{2,3}(0)=1$, $b_{2,3}(0)=c_{2,3}(n)=d_{2,3}(n)=0$ and $f_{2,3}(0)=6$. We wrote a computer program to obtain the recursion relations which are given in the appendix.
Some values of $f_{2,3}(n)$, $a_{2,3}(n)$, $b_{2,3}(n)$, $c_{2,3}(n)$, $d_{2,3}(n)$ are listed in Table \ref{tablesg23}. These numbers grow exponentially, and do not have simple integer factorizations.

\bigskip

\begin{table}[htbp]
\caption{\label{tablesg23} The first few values of $f_{2,3}(n)$, $a_{2,3}(n)$, $b_{2,3}(n)$, $c_{2,3}(n)$, $d_{2,3}(n)$.}
\begin{center}
\begin{tabular}{|c||r|r|r|}
\hline\hline 
$n$          & 0 &      1 &                                  2 \\ \hline\hline 
$f_{2,3}(n)$ & 6 & 19,602 & 55,220,940,611,523,034,547,131,584 \\ \hline 
$a_{2,3}(n)$ & 1 &    140 &        272,601,409,439,732,172,800 \\ \hline 
$b_{2,3}(n)$ & 0 &    918 &     17,241,894,275,103,011,071,296 \\ \hline 
$c_{2,3}(n)$ & 0 &    966 &    293,676,957,591,553,508,446,272 \\ \hline
$d_{2,3}(n)$ & 0 &  7,458 & 53,353,791,891,866,457,036,989,376 \\ \hline\hline 
\end{tabular}
\end{center}
\end{table}

\bigskip

\begin{lemma} \label{lemmasg23b} The asymptotic growth constant for the number of acyclic orientations on $SG_{2,3}(n)$ is bounded:
\beq
\frac{5}{7\times 6^m} \ln d_{2,3}(m) < z_{SG_{2,3}} < \frac{5}{7\times 6^m} \ln f_{2,3}(m)  \ ,
\label{zsg23}
\eeq
where $m$ is a positive integer.
\end{lemma}

It is clear that $a_{2,3}(n) \ll b_{2,3}(n) \ll c_{2,3}(n) \ll d_{2,3}(n)$ for large $n$. As for the ordinary Sierpinski gasket, $a_{2,3}(n)$, $b_{2,3}(n)$, $d_{2,3}(n)$ are negligible compared with $d_{2,3}(n)$ such that $f_{2,3}(n) \sim d_{2,3}(n)$ for large $n$. By Eqs. (\ref{f23eq}) and (\ref{d23eq}) in the appendix, we have the upper and lower bounds for $f_{2,3}(n)$:
\beq
d_{2,3}^6(n-1) < d_{2,3}(n) < f_{2,3}(n) < f_{2,3}^6(n-1) \ ,
\eeq
such that
\beq
d_{2,3}(m)^{6^{n-m}} < f_{2,3}(n) < f_{2,3}(m)^{6^{n-m}} \ ,
\eeq
where $m$ is a fixed integer smaller than $n$. With the definition for $z_{SG_{2,3}}$ given in Eq. (\ref{zdef}) and the vertex number of $SG_{2,3}(n)$ by Eq. (\ref{vsg23}), Eq. (\ref{zsg23}) is established.
We have the following proposition.

\bigskip

\begin{propo} \label{proposg23} The asymptotic growth constant for the number of acyclic orientations on the two-dimensional Sierpinski gasket $SG_{2,3}(n)$ in the large $n$ limit is $z_{SG_{2,3}}=1.176059211520985...$.

\end{propo}

\bigskip

The convergence of the upper and lower bounds remains rapid. When $m$ is as small as two, the difference between the upper bound in Eq. (\ref{zsg23}) and $z_{SG_{2,3}}$ is about $4 \times 10^{-9}$ by the values given in Table \ref{tablesg23}. More than a hundred significant figures for $z_{SG_{2,3}}$ can be obtained when $m$ is equal to six.

\section{Upper Bounds of the asymptotic growth constants}
\label{upperbounds}

By a similar argument as in Lemma \ref{lemmasg2b} using Eq. (\ref{v}), we have the upper bound of the asymptotic growth constant for the number of acyclic orientations on $SG_d(n)$:
\beq
z_{SG_d} < \frac{2}{(d+1)^{m+1}} \ln N_{AO}(SG_d(m)) \equiv \bar z_{SG_d}(m)  \ ,
\label{zsgd}
\eeq
with $m$ a positive integer. To see how fast the convergence of the upper bound to the true value, we list the first few values of $\bar z_{SG_2}(m)$ and the ratio of $z_{SG_2}$ to $\bar z_{SG_2}(m)$ in Table \ref{zsg2table}. The number of acyclic orientations is also calculated for the three-dimensional Sierpinski gasket $SG_3(n)$ for $0 \le n \le 4$, and the upper bound $\bar z_{SG_3}(m)$ is given in Table \ref{zsg2table}. 

\bigskip

\begin{table}
\caption{\label{zsg2table} Numerical values of $\bar z_{SG_2}(m)$, the ratio of $z_{SG_2}$ to $\bar z_{SG_2}(m)$, and $\bar z_{SG_3}(m)$. }
\begin{center}
\begin{tabular}{|c|c|c|c|}
\hline\hline 
$m$ & $\bar z_{SG_2}(m)$ & $z_{SG_2}/\bar z_{SG_2}(m)$ & $\bar z_{SG_3}(m)$ \\ \hline\hline 
0   &  1.194506312818703 & 0.9437363858517446 & 1.589026915173972 \\ \hline
1   &  1.130576963384974 & 0.9971006902187856 & 1.449952098843616 \\ \hline
2   &  1.127331566378142 & 0.9999711745483802 & 1.442835469614084 \\ \hline
3   &  1.127299079278937 & 0.9999999922448969 & 1.442740614008092 \\ \hline
4   &  1.127299070536618 & 0.9999999999999985 & 1.442740560266077 \\ \hline
5   &  1.127299070536616 & 0.9999999999999999 & - \\ \hline
\hline 
\end{tabular}
\end{center}
\end{table}

Although the number $N_{AO}(SG_d(m))$ for general $m$ is difficult to obtain, it is known for $m=0$ and arbitrary $d$. We first recall that $SG_d(0)$ at stage zero is a complete graph with $(d+1)$ vertices, each of which is adjacent to all of the other vertices. As the chromatic polynomial for the complete graph with $(d+1)$ vertices is given by $P(SG_d(0),q) = q(q-1)...(q-d)$, the number of acyclic orientations on $SG_d(0)$ is $(d+1)!$. Therefore, we have
\beq
\bar z_{SG_d}(0) = \frac{2}{d+1} \ln (d+1)! \ .
\eeq
The first few values of the upper bound $\bar z_{SG_d}(0)$ is given in Table \ref{zsgdtable}. 

\bigskip

\begin{table}
\caption{\label{zsgdtable} Fractal dimension $D$ for $SG_d$ and numerical values of $\bar z_{SG_d}(0)$ for $2 \le d \le 10$. }
\begin{center}
\begin{tabular}{|c|c|c|}
\hline\hline 
$d$ &  $D$      & $\bar z_{SG_d}(0)$ \\ \hline\hline 
2   & 1.5849625 &  1.194506312818703 \\ \hline
3   & 2         &  1.589026915173972 \\ \hline
4   & 2.3219280 &  1.914996697112818 \\ \hline
5   & 2.5849625 &  2.193083737336700 \\ \hline
6   & 2.8073549 &  2.435760388875832 \\ \hline
7   & 3         &  2.651150725686312 \\ \hline
8   & 3.1699250 &  2.844850551129215 \\ \hline
9   & 3.3219280 &  3.020882514615103 \\ \hline
10  & 3.4594316 &  3.182237790158888 \\ \hline\hline 
\end{tabular}
\end{center}
\end{table}

\bigskip

For the generalized Sierpinski gasket $SG_{2,b}(n)$ with dimension equal to two, the number of vertices can be calculated to be
\beq
v(SG_{2,b}(n)) = \frac{b+4}{b+2} \left[ \frac{b(b+1)}{2} \right ]^n + \frac{2(b+1)}{b+2} \ .
\eeq
The upper bound of the asymptotic growth constant for the number of acyclic orientations on $SG_{2,b}(n)$ is given by
\beq
z_{SG_{2,b}} < \left ( \frac{b+2}{b+4} \right ) \frac{\ln N_{AO}(SG_{2,b}(m))}{\left[b(b+1)/2 \right]^m} \equiv \bar z_{SG_{2,b}}(m)  \ ,
\label{zsg2b}
\eeq
with $m$ a positive integer. To see how fast the convergence of the upper bound to the true value, we list the first few values of $\bar z_{SG_{2,3}}(m)$ and the ratio of $z_{SG_{2,3}}$ to $\bar z_{SG_{2,3}}(m)$ in Table \ref{zsg23table}. Notice that the convergence of $z_{SG_{2,3}}$ to $\bar z_{SG_{2,3}}(m)$ is faster than that of $z_{SG_2}$ to $\bar z_{SG_2}(m)$. The number of acyclic orientations is also calculated for the generalized Sierpinski gasket $SG_{2,4}(n)$ for $0 \le n \le 2$, and the upper bound $\bar z_{SG_{2,4}}(m)$ is given in Table \ref{zsg23table}. 

\bigskip

\begin{table}
\caption{\label{zsg23table} Numerical values of $\bar z_{SG_{2,3}}(m)$, the ratio of $z_{SG_{2,3}}$ to $\bar z_{SG_{2,3}}(m)$, and $\bar z_{SG_{2,4}}(m)$. }
\begin{center}
\begin{tabular}{|c|c|c|c|}
\hline\hline 
$m$ & $\bar z_{SG_{2,3}}(m)$ & $z_{SG_{2,3}}/\bar z_{SG_{2,3}}(m)$ & $\bar z_{SG_{2,4}}(m)$ \\ \hline\hline 
0   &  1.279828192305753 & 0.9189196007646803 & 1.343819601921041 \\ \hline
1   &  1.176593676289181 & 0.9995457524726109 & 1.213368082437441 \\ \hline
2   &  1.176059215716391 & 0.9999999964326572 & 1.213273099891158 \\ \hline
3   &  1.176059211520985 & 0.9999999999999999 & - \\ \hline
\hline 
\end{tabular}
\end{center}
\end{table}

Although the number $N_{AO}(SG_{2,b}(m))$ for general $m$ is difficult to obtain, it is always equal to six for stage zero since $SG_{2,b}(0)$ is the equilateral triangle for any $b$. Therefore, we have
\beq
\bar z_{SG_{2,b}}(0) = \left ( \frac{b+2}{b+4} \right ) \ln 6  \ .
\label{zsg2bn}
\eeq
We list the first few values of $\bar z_{SG_{2,b}}(0)$ in Table \ref{zsg2btable}. 

\bigskip

\begin{table}
\caption{\label{zsg2btable} Fractal dimension $D$ for $SG_{2,b}$ and numerical values of $\bar z_{SG_{2,b}}(0)$ for $3 \le b \le 10$ and $b \to \infty$.}
\begin{center}
\begin{tabular}{|c|c|c|}
\hline\hline 
$b$      &  $D$      & $\bar z_{SG_{2,b}}$ \\ \hline\hline 
3        & 1.6309297 & 1.279828192305753   \\ \hline
4        & 1.6609640 & 1.343819601921041   \\ \hline
5        & 1.6826061 & 1.393590698288487   \\ \hline
6        & 1.6991803 & 1.433407575382444   \\ \hline
7        & 1.7124143 & 1.465985020277499   \\ \hline
8        & 1.7233083 & 1.493132891023379   \\ \hline
9        & 1.7324867 & 1.516104166269892   \\ \hline
10       & 1.7403626 & 1.535793830766904   \\ \hline
$\infty$ & 2         & 1.791759469228055   \\ \hline\hline 
\end{tabular}
\end{center}
\end{table}

\bigskip

\acknowledgments
This research was partially supported by the NSC grant NSC-97-2112-M-006-007-MY3. 

\bigskip

\appendix

\section{Recursion relations for $SG_{2,3}(n)$}

We give the recursion relations for the generalized Sierpinski gasket $SG_{2,3}(n)$ here. Since the subscript is $(d,b)=(2,3)$ for all the quantities throughout this appendix, we will use the simplified notation $f_{n+1}$ to denote $f_{2,3}(n+1)$ and similar notations for other quantities. For any non-negative integer $n$, we have
\beqs
f_{n+1} & = & f_n^6 - 2 (3a_n+2b_n+c_n)^3 (501a_n^3 + 1638a_n^2b_n + 1773a_n^2c_n + 297a_n^2d_n + 1755a_nb_n^2 \cr\cr
& & + 3744a_nb_nc_n + 621a_nb_nd_n + 1971a_nc_n^2 + 648a_nc_nd_n + 54a_nd_n^2 + 614b_n^3 + 1929b_n^2c_n \cr\cr
& & + 318b_n^2d_n + 1986b_nc_n^2 + 651b_nc_nd_n + 54b_nd_n^2 + 667c_n^3 + 327c_n^2d_n + 54c_nd_n^2 + 3d_n^3) \ , \cr & &
\label{f23eq}
\eeqs
\beqs
a_{n+1} & = & 140a_n^6 + 644a_n^5b_n + 588a_n^5c_n + 45a_n^5d_n + 1141a_n^4b_n^2 + 1988a_n^4b_nc_n + 132a_n^4b_nd_n \cr\cr
& & + 812a_n^4c_n^2 + 84a_n^4c_nd_n + 978a_n^3b_n^3 + 2445a_n^3b_n^2c_n + 143a_n^3b_n^2d_n + 1908a_n^3b_nc_n^2 \cr\cr
& & + 176a_n^3b_nc_nd_n + 460a_n^3c_n^3 + 50a_n^3c_n^2d_n + 407a_n^2b_n^4 + 1300a_n^2b_n^3c_n + 68a_n^2b_n^3d_n \cr\cr
& & + 1455a_n^2b_n^2c_n^2 + 122a_n^2b_n^2c_nd_n + 668a_n^2b_nc_n^3 + 68a_n^2b_nc_n^2d_n + 104a_n^2c_n^4 + 12a_n^2c_n^3d_n \cr\cr
& & + 66a_nb_n^5 + 253a_nb_n^4c_n + 12a_nb_n^4d_n + 362a_nb_n^3c_n^2 + 28a_nb_n^3c_nd_n + 239a_nb_n^2c_n^3 \cr\cr
& & + 23a_nb_n^2c_n^2d_n + 72a_nb_nc_n^4 + 8a_nb_nc_n^3d_n + 8a_nc_n^5 + a_nc_n^4d_n \ ,
\label{a23eq}
\eeqs
\beqs
b_{n+1} & = & 918a_n^6 + 5360a_n^5b_n + 4932a_n^5c_n + 504a_n^5d_n + 12403a_n^4b_n^2 + 21524a_n^4b_nc_n + 1905a_n^4b_nd_n \cr\cr
& & + 8606a_n^4c_n^2 + 1200a_n^4c_nd_n + 14634a_n^3b_n^3 + 35910a_n^3b_n^2c_n + 2786a_n^3b_n^2d_n + 26784a_n^3b_nc_n^2 \cr\cr
& & + 3260a_n^3b_nc_nd_n + 5840a_n^3c_n^3 + 776a_n^3c_n^2d_n + 9338a_n^2b_n^4 + 28846a_n^2b_n^3c_n + 1988a_n^2b_n^3d_n \cr\cr
& & + 30228a_n^2b_n^2c_n^2 + 3284a_n^2b_n^2c_nd_n + 12248a_n^2b_nc_n^3 + 1502a_n^2b_nc_n^2d_n + 1514a_n^2c_n^4 \cr\cr
& & + 192a_n^2c_n^3d_n + 3072a_nb_n^5 + 11230a_nb_n^4c_n + 696a_nb_n^4d_n + 14774a_nb_n^3c_n^2 \cr\cr
& & + 1456a_nb_n^3c_nd_n + 8426a_nb_n^2c_n^3 + 962a_nb_n^2c_n^2d_n + 1968a_nb_nc_n^4 + 236a_nb_nc_n^3d_n \cr\cr
& & + 140a_nc_n^5 + 16a_nc_n^4d_n + 409b_n^6 + 1704b_n^5c_n + 96b_n^5d_n + 2652b_n^4c_n^2 + 240b_n^4c_nd_n \cr\cr
& & + 1906b_n^3c_n^3 + 204b_n^3c_n^2d_n + 633b_n^2c_n^4 + 72b_n^2c_n^3d_n + 84b_nc_n^5 + 9b_nc_n^4d_n + 2c_n^6 \ ,
\label{b23eq}
\eeqs
\beqs
c_{n+1} & = & 966a_n^6 + 6848a_n^5b_n + 8040a_n^5c_n + 1584a_n^5d_n + 19150a_n^4b_n^2 + 43004a_n^4b_nc_n \cr\cr
& & + 7746a_n^4b_nd_n + 23198a_n^4c_n^2 + 7617a_n^4c_nd_n + 513a_n^4d_n^2 + 27096a_n^3b_n^3 + 87117a_n^3b_n^2c_n \cr\cr
& & + 14345a_n^3b_n^2d_n + 89220a_n^3b_nc_n^2 + 26528a_n^3b_nc_nd_n + 1602a_n^3b_nd_n^2 + 28820a_n^3c_n^3 \cr\cr
& & + 11336a_n^3c_n^2d_n + 1152a_n^3c_nd_n^2 + 27a_n^3d_n^3 + 20540a_n^2b_n^4 + 84010a_n^2b_n^3c_n + 12692a_n^2b_n^3d_n \cr\cr
& & + 122577a_n^2b_n^2c_n^2 + 33260a_n^2b_n^2c_nd_n + 1830a_n^2b_n^2d_n^2 + 74852a_n^2b_nc_n^3 + 26816a_n^2b_nc_n^2d_n \cr\cr
& & + 2514a_n^2b_nc_nd_n^2 + 54a_n^2b_nd_n^3 + 15926a_n^2c_n^4 + 6558a_n^2c_n^3d_n + 786a_n^2c_n^2d_n^2 + 27a_n^2c_nd_n^3 \cr\cr
& & + 7950a_nb_n^5 + 38827a_nb_n^4c_n + 5412a_nb_n^4d_n + 71948a_nb_n^3c_n^2 + 17980a_nb_n^3c_nd_n \cr\cr
& & + 912a_nb_n^3d_n^2 + 62663a_nb_n^2c_n^3 + 20741a_nb_n^2c_n^2d_n + 1812a_nb_n^2c_nd_n^2 + 36a_nb_n^2d_n^3 \cr\cr
& & + 25404a_nb_nc_n^4 + 9800a_nb_nc_n^3d_n + 1110a_nb_nc_n^2d_n^2 + 36a_nb_nc_nd_n^3 + 3812a_nc_n^5 \cr\cr
& & + 1624a_nc_n^4d_n + 216a_nc_n^3d_n^2 + 9a_nc_n^2d_n^3 + 1234b_n^6 + 6924b_n^5c_n + 896b_n^5d_n + 15333b_n^4c_n^2 \cr\cr
& & + 3560b_n^4c_nd_n + 168b_n^4d_n^2 + 17026b_n^3c_n^3 + 5264b_n^3c_n^2d_n + 432b_n^3c_nd_n^2 + 8b_n^3d_n^3 + 9933b_n^2c_n^4 \cr\cr
& & + 3622b_n^2c_n^3d_n + 390b_n^2c_n^2d_n^2 + 12b_n^2c_nd_n^3 + 2880b_nc_n^5 + 1174b_nc_n^4d_n + 150b_nc_n^3d_n^2 \cr\cr
& & + 6b_nc_n^2d_n^3 + 326c_n^6 + 145c_n^5d_n + 21c_n^4d_n^2 + c_n^3d_n^3 \ ,
\label{c23eq}
\eeqs
\beqs
d_{n+1} & = & 7458a_n^6 + 60264a_n^5b_n + 75780a_n^5c_n + 17820a_n^5d_n + 195930a_n^4b_n^2 + 482400a_n^4b_nc_n \cr\cr
& & + 108972a_n^4b_nd_n + 292950a_n^4c_n^2 + 128844a_n^4c_nd_n + 13446a_n^4d_n^2 + 328224a_n^3b_n^3 \cr\cr
& & + 1185624a_n^3b_n^2c_n + 257292a_n^3b_n^2d_n + 1406448a_n^3b_nc_n^2 + 593280a_n^3b_nc_nd_n \cr\cr
& & + 59400a_n^3b_nd_n^2 + 547032a_n^3c_n^3 + 335592a_n^3c_n^2d_n + 65016a_n^3c_nd_n^2 + 3996a_n^3d_n^3 \cr\cr
& & + 299250a_n^2b_n^4 + 1409328a_n^2b_n^3c_n + 294264a_n^2b_n^3d_n + 2449548a_n^2b_n^2c_n^2 + 993708a_n^2b_n^2c_nd_n \cr\cr
& & + 95940a_n^2b_n^2d_n^2 + 1859616a_n^2b_nc_n^3 + 1097496a_n^2b_nc_n^2d_n + 205560a_n^2b_nc_nd_n^2 \cr\cr
& & + 12312a_n^2b_nd_n^3 + 519534a_n^2c_n^4 + 396072a_n^2c_n^3d_n + 108036a_n^2c_n^2d_n^2 + 12636a_n^2c_nd_n^3 \cr\cr
& & + 540a_n^2d_n^4 + 141120a_nb_n^5 + 812700a_nb_n^4c_n + 163728a_nb_n^4d_n + 1841472a_nb_n^3c_n^2 \cr\cr
& & + 721296a_nb_n^3c_nd_n + 67536a_nb_n^3d_n^2 + 2049912a_nb_n^2c_n^3 + 1170180a_nb_n^2c_n^2d_n \cr\cr
& & + 213336a_nb_n^2c_nd_n^2 + 12528a_nb_n^2d_n^3 + 1120104a_nb_nc_n^4 + 828576a_nb_nc_n^3d_n \cr\cr
& & + 221112a_nb_nc_n^2d_n^2 + 25488a_nb_nc_nd_n^3 + 1080a_nb_nd_n^4 + 240228a_nc_n^5 + 216252a_nc_n^4d_n \cr\cr
& & + 75384a_nc_n^3d_n^2 + 12852a_nc_n^2d_n^3 + 1080a_nc_nd_n^4 + 36a_nd_n^5 + 26974b_n^6 + 182568b_n^5c_n \cr\cr
& & + 35616b_n^5d_n + 506490b_n^4c_n^2 + 192432b_n^4c_nd_n + 17568b_n^4d_n^2 + 736816b_n^3c_n^3 \cr\cr
& & + 409080b_n^3c_n^2d_n + 73008b_n^3c_nd_n^2 + 4224b_n^3d_n^3 + 592746b_n^2c_n^4 + 428100b_n^2c_n^3d_n \cr\cr
& & + 112356b_n^2c_n^2d_n^2 + 12816b_n^2c_nd_n^3 + 540b_n^2d_n^4 + 250176b_nc_n^5 + 220956b_nc_n^4d_n \cr\cr
& & + 76104b_nc_n^3d_n^2 + 12888b_nc_n^2d_n^3 + 1080b_nc_nd_n^4 + 36b_nd_n^5 + 43354c_n^6 + 45132c_n^5d_n \cr\cr
& & + 19206c_n^4d_n^2 + 4308c_n^3d_n^3 + 540c_n^2d_n^4 + 36c_nd_n^5 + d_n^6 \ .
\label{d23eq}
\eeqs

\bibliographystyle{abbrvnat}

\begin{thebibliography}{1}

\bibitem[Bollob\'as and Szab\'o(1998)]{bollobas} 
B.~Bollob\'as and T.~Szab\'o, The oriented cycle game, \emph{Discrete Math.}, 186: 55-67, 1998.

\bibitem[Reidys(1998)]{reidys} 
C.~M.~Reidys, Acyclic orientations of random graphs, \emph{Adv. Appl. Math.}, 21: 181-192, 1998.

\bibitem[Gebhard and Sagan(2000)]{gebhard} 
D.~D.~Gebhard and B.~E.~Sagan, Sinks in acyclic orientations of graphs, \emph{J. Comb. Theory B}, 80: 130-146, 2000.

\bibitem[Gioan and Las Vergnas(2005)]{gioan} 
E.~Gioan and M.~Las Vergnas, Activity preserving bijections between spanning trees and orientations in graphs, \emph{Discrete Math.}, 298: 169-188, 2005.

\bibitem[Kr\'{a}lovi\v{c} and Ru\v{z}i\v{c}ka(2007)]{kralovic} 
R.~Kr\'{a}lovi\v{c} and P.~Ru\v{z}i\v{c}ka, Ranks of graphs: The size of acyclic orientation cover for deadlock-free packet routing, \emph{Theor. Comput. Sci.}, 374: 203-213, 2007.

\bibitem[Arantes et~al.(2009)]{arantes} 
G.~M.~Arantes Jr., F.~M.~G.~Fran\c{c}a and C.~A.~Martinhon, Randomized generation of acyclic orientations upon anonymous distributed systems, \emph{J. Parallel Distrib. Comput.}, 69: 239-246, 2009.

\bibitem[Welsh(1993)]{welsh}
D.~J.~A.~Welsh, \emph{Complexity: Knots, Colourings, and Counting (London Math. Soc. Lecture Notes series 186)}, Cambridge University Press, Cambridge, 1993.

\bibitem[Stanley(1973)]{stanley} 
R.~P.~Stanley, Acyclic orientations of graphs, \emph{Discrete Math.}, 5: 171-178, 1973.

\bibitem[Mandelbrot(1982)]{mandelbrot}
B.~B.~Mandelbrot, \emph{The Fractal Geometry of Nature}, Freeman, San Francisco, 1982.

\bibitem[Falconer(2003)]{Falconer}
K.~J.~Falconer, \emph{Fractal Geometry: Mathematical Foundations and Applications}, 2nd ed., Wiley, Chichester, 2003.

\bibitem[Biggs(1993)]{bbook}
N.~L.~Biggs, \emph{Algebraic Graph Theory}, 2nd ed., Cambridge University Press, Cambridge, 1993.

\bibitem[Harary(1969)]{fh} 
F.~Harary, \emph{Graph Theory}, Addison-Wesley, New York, 1969.

\bibitem[Burton and Pemantle(1993)]{burton93} 
R.~Burton and R.~Pemantle, Local characteristics, entropy and limit theorems for spanning trees and domino tilings via transfer-impedances, \emph{Ann. Probab.}, 21: 1329-1371, 1993.

\bibitem[Gefen and Aharony(1981)]{Gefen81}
Y.~Gefen and A.~Aharony, Solvable fractal family, and its possible relation to the backbone at percolation, \emph{Phys. Rev. Lett.}, 47: 1771-1774, 1981.

\bibitem[Hilfer and Blumen(1984)]{Hilfer}
R.~Hilfer and A.~Blumen, Renormalisation on Sierpinski-type fractals, \emph{J. Phys. A: Math. Gen.}, 17: L537-L545, 1984.


\end{thebibliography}

\end{document}